\documentclass[12pt]{iopart}

\usepackage{graphicx}
\usepackage{subfigure}
\usepackage{epstopdf}
\usepackage{setspace}
\usepackage[normalem]{ulem}
\usepackage[pdftex]{hyperref}
\hypersetup{colorlinks=true,citecolor=blue}

\begin{document}

\title{Tight packing of a flexible rod in two-dimensional cavities.}

\author{T A Sobral$^1$, M A F Gomes $^1$}

\address{$^1$ Departamento de F\'isica, Universidade Federal de Pernambuco, 50670-901, Recife, PE, Brasil.}

\eads{\mailto{mafg@ufpe.br}} 

\begin{abstract}
The present work deals with the injection and packing of a flexible polymeric rod of length $L$ into a simply connected rectangular domain of area $XY$. As the injection proceeds, the rod bends over itself and it stores elastic energy in closed loops. In a typical experiment $N$ of these loops can be identified inside the cavity in the jammed state. We have performed an extensive experimental analysis of the total length $L(N, X, Y)$ in the tight packing limit, and have obtained robust power laws relating these variables. Additionally, we have examined a version of this packing problem when the simply connected domain is partially occupied with free discs of fixed size. The experimental results were obtained with 27 types of cavities and obey a single equation of state valid for the tight packing of rods in domains of different topologies. Besides its intrinsic theoretical interest and generality, the problem examined here could be of interest in a number of studies including package models of DNA and polymers in several complex environments.
\end{abstract}



\submitto{\JPD}

\maketitle

\hypersetup{linkcolor=blue}


\section{Introduction}
\label{secI}

Physics in less than three dimensions is a subject of wide theoretical and applied interest in the last decades~\cite{gumbs11,lerner02}. On the other hand, the packing properties of atoms, molecules, polymers, and other components is an essential issue in defining the full physical behaviour of materials in general~\cite{weaire08}. As examples, the influence of the packing fraction on the physical properties of systems has recently been studied for a number of different situations, including the thermal and magnetic properties in Ni nanorods in several matrices~\cite{piraux13}; the mechanical and ergodic properties of granular matter~\cite{zhao14}; the collective motion of the clusters in colonies of bacteria~\cite{peruani12}; the rheometry in shear thickening suspensions~\cite{brown09}; the drag forces in loose granular media~\cite{bruyn04}; and the propensity to crystallization~\cite{luchnikov02}, among many others.

In the past, Evans and Ferrar~\cite{evans89} reported in this Journal a study of the packing of rods with aspect ratio between 1 and 30, a subject of importance for many technological areas, including the techniques of moulding of short fibre reinforced composites, and the packing of high-aspect-ratio ceramic particles, as well as liquid crystals. Here we are interested in the study of the formation and organization of structures that grow within a fixed domain when a single rod is constrained to be accommodated in a two-dimensional transparent cavity in the jamming limit tending to maximize the coverage fraction~\cite{donato02,stoop08,bayart11}. The rod has a length $L$ much greater than its diameter, and the cavity has a height that only allows structures with a single layer of the material. In these circumstances, the morphologies of the structures formed have a larger surface area as compared to the bulk content, and they can be easily accessed for image recording. Beyond the intrinsic theoretical interests, the problem discussed here has connections with some important issues in the life sciences, as the packing of DNA molecules in viral capsids~\cite{purohit05,smith11,maycon08,katzav06}, the physics of an unbranched polymer in confined geometries~\cite{sakaue06,gennes79}, and the problem of packing and delivery of a polymer-like drug with the aid of nano-capsids~\cite{Schmaljohann06,solovev12}. The packing process studied here generates a planar pattern with a set of loops that is somewhat similar to the classical two-dimensional packing of discs, a problem that has a wide range of applications in technology and science, including the structures of glasses, crystals, granular piles, molecular films and foams~\cite{donato03}. 

Besides the study of the packing of an elastic rod in a simply connected domain, additional studies considering the packing of the rod in the presence of a given number of fixed pins were also reported~\cite{gomes10, gomes13}. These non-simply connected domains have a very strong influence and are responsible for the decrease of the total length of the rod in the jammed state.  Two effects appear in these situations: a reduction of the available area for the rod, and a change in the topology of the cavity itself. In the present study the contribution due exclusively to the reduction of the available area $A$ is investigated as well as the contribution of the changes in topology by the presence of a number of free discs. The present paper, to the best of our knowledge, is the first study of two-dimensional packing of a rod using free discs. This novelty is somewhat reminiscent of polymer and biological contexts,  e.g. in the study of the statistical configurations of a long unbranched polymer in the presence of a dispersal of large (substrate) molecules, as in enzymatic catalysis, and in the compaction of DNA in chromosomes in the presence of histones and a gamut of similar situations~\cite{bakajin98,turner02,yu14}.

The paper is organized as follows: In section~\ref{secED}, there is a description of the apparatus used to perform all the experiments involved in this study. The results and discussions are in section~\ref{secRD} distributed in three subsections: in section~\ref{secCWO} we analyze the data associated with cavities without discs; in section~\ref{secCO} we analyze the competition between the packing of circular discs and the packing of the rod; and in section~\ref{secSM} some models and further comments are made. The main conclusions and perspectives are made in section~\ref{secCP}.


\section{Experimental details}
\label{secED}

\begin{figure}[!hb]
\centering
\includegraphics[width=.6\linewidth]{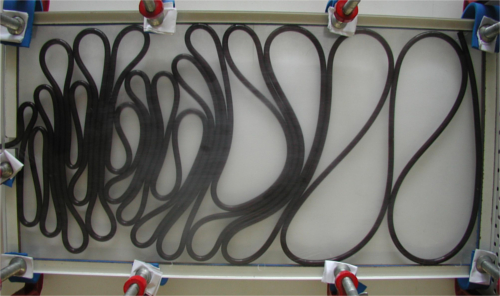}
\caption{Tight packing of a polymeric rod in a rectangular cavity of $X=40$~cm and $Y=20$ cm. In this particular case it can be seen $N=34$ loops corresponding to a total length $L=589$~cm.}
\label{fig1}
\end{figure}

In the present study the cavities are of rectangular shape where the depth $X$ and the width $Y$ are controlled parameters. These cavities consist of two 22~cm $\times$ 42~cm $\times$ 1.0~cm  planar acrylic plates separated by wooden spacers of 0.6~cm $\times$ 1.0~cm with several lengths in order to obtain different available areas with a range of aspect ratios. The cavity is held rigid by a set of fasteners on its boundaries as shown in Fig.~\ref{fig1}. The polymeric rod used is a hollow cylinder with outer diameter of $\zeta = 0.50$~cm and inner diameter of $\zeta_2 = 0.25$~cm which is adequate to make a flexible rod that does not present wrinkles. The cable is made of flexible polyvinyl chloride and presents flexural rigidity of $1.6$ x $10^{-3}$~Nm$^2$ and torsional rigidity of $1.1$ x $10^{-3}$~Nm$^2$. In this study the rod is manually inserted inside the cavity at an approximated ratio of 1~cm/s through an unique injection channel composed by a spatial gap of $\approx 1$ cm located in the centre of one of the smaller edges.

In the beginning of the injection, the tip of the rod advances until it touches the opposite extremity of the cavity, when the rod bends in order to occupy the available area of the cavity. Frictional effects prevents the free end of the rod from easily slide along the box contour, therefore the rod does not assume the configuration of minimum elastic energy. The system studied here is intrinsically non-thermal because the thermal energy is much smaller than the relevant typical energies associated with the dynamics of the packing. The first self-contact of the rod encloses what it is called loop: a two-dimensional shape that has a bulge in one extremity and a tip in the other. With the progress of the injection new self-contacts as well as new loops are formed, while the injection of the rod becomes progressively harder. This is the description of a jamming process, where the system becomes rigid as its density increases. Similar jammed systems are glasses, emulsions, pastes, gels, and sand~\cite{liu98}. These materials exhibit a jamming transition that come to attention in recent years~\cite{biroli07}. In our experiments, the insertion finishes when the system reaches the jammed state, i.e. when further injection is impossible to occur. After that, the total quantity of loops $N$ is counted and the total length of the rod $L$ is taken with a measuring tape. Each experiment of packing of rod in a cavity is performed 10 times within identical initial conditions. A typical configuration of tight packing is shown in Fig.~\ref{fig1}: a cascade of loops that are smaller and numerous near the injection channel and follow a gradient of sizes. All experiments are performed in a dry regime, free of lubricants. Additionally, the plastic rod was reused during all experiments without presenting fatigue, with an appropriate treatment of cleaning and unfolding after each use. In the following, each set of cavities used in the experiments is discussed separately.

\subsection{Cavities with fixed aspect ratio $B=2$}

\begin{figure}[b]
\centering
\includegraphics[height=0.30\linewidth]{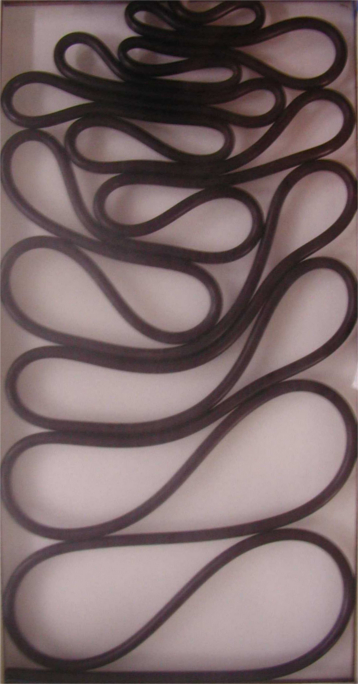}
\includegraphics[height=0.30\linewidth]{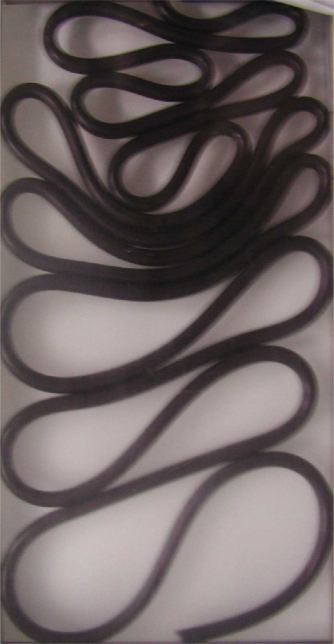}
\includegraphics[height=0.30\linewidth]{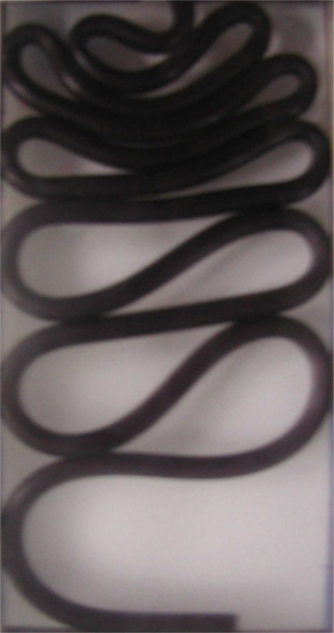}
\includegraphics[height=0.30\linewidth]{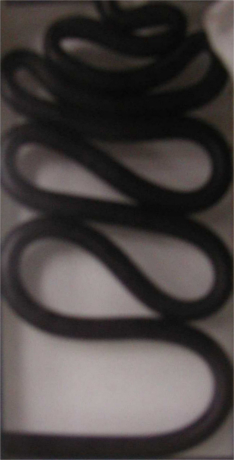}
\includegraphics[height=0.30\linewidth]{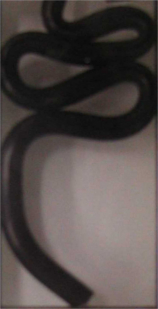}
\includegraphics[height=0.30\linewidth]{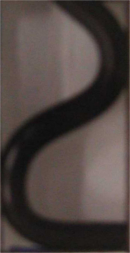}
\caption{Tight packing of a rod in cavities of area $A=\{392, 200, 84.5, 40.5, 18, 8\}$ cm$^2$ (from left to right), and a fixed aspect ratio of $B= X / Y = 2$ (rescaled).}
\label{fig2}
\end{figure}

In order to vary the area $A = XY$ of the cavity without varying the aspect ratio $B = XY^{-1}$, spacers were used corresponding to cavities of  $Y = \{2, 3, 4.5, 6.5, 10, 14,20 \} $ cm and the chosen was $B=2$ which corresponds to $A = \{8, 18, 40.5, 84.5, 200, 392, 800\} $ cm$^2$ (two decades of variability in the available area). A typical configuration of jamming of the rod is shown in Fig.~\ref{fig1} for the biggest cavity ($A=800$ cm$^2$). Figure~\ref{fig2} shows some typical configurations obtained with other cavities. Due to the zoom made in order to fit the cavity, there is an apparent effect of enlarging the diameter of the rod.

The configurations for tight packing of the rod in rectangular cavities present more directional order than the patterns reported in literature for circular cavities \cite{donato02, stoop08, maycon08}, because the cavity does not allow rotational motion easily. As the area increases the geometric patterns exhibited by the rods present many loops that do not reach the boundaries of the cavity, an ingredient that increases the complexity of the pattern. For cavities of intermediary sizes, a larger fraction of the loops touches the perimeter of the cavity. Finally, for the smallest cavity the formation of loops is not observed at all.

\subsection{Cavities with fixed depth $X=40$ cm}

\begin{figure}[!ht]
\centering
\includegraphics[height=0.30\linewidth]{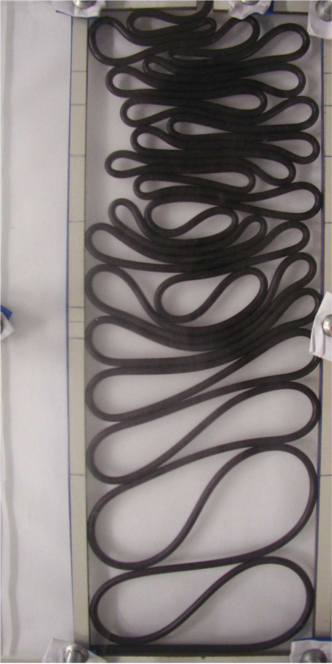}
\includegraphics[height=0.30\linewidth]{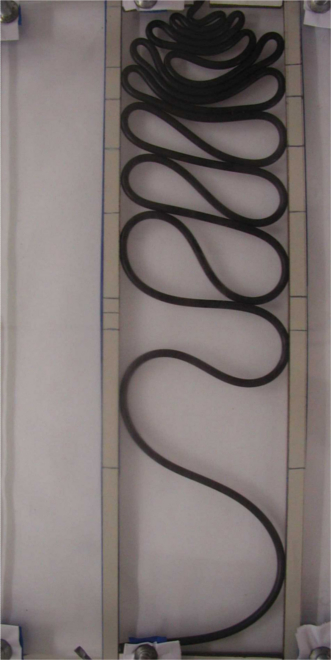}
\includegraphics[height=0.30\linewidth]{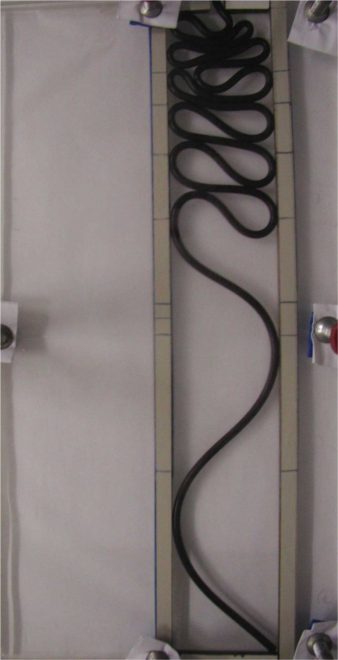}
\includegraphics[height=0.30\linewidth]{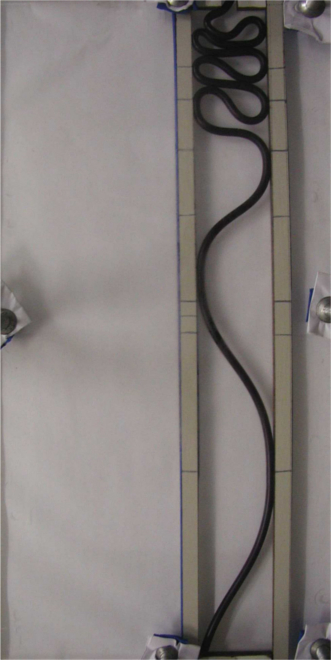}
\includegraphics[height=0.30\linewidth]{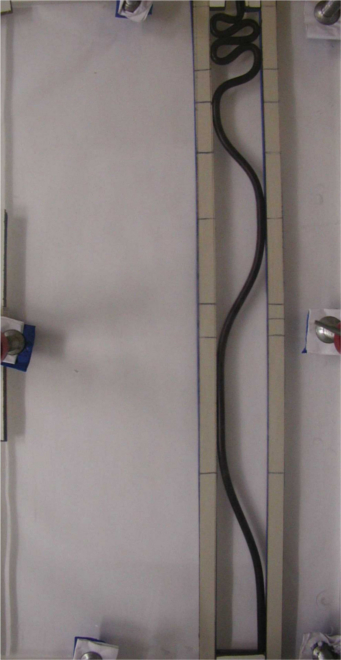}
\includegraphics[height=0.30\linewidth]{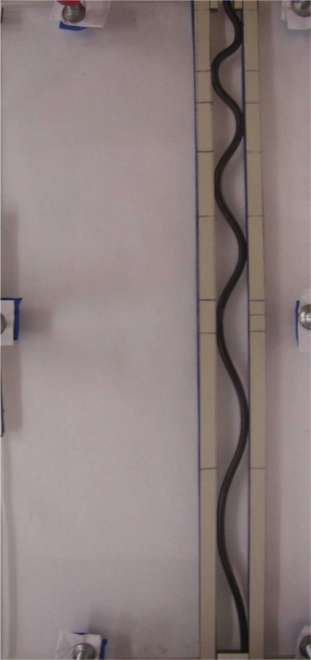}
\caption{Tight packing of a rod in cavities of width $Y = \{14, 10, 6.5, 4.5, 3, 2\}$ cm (from left to right), and a fixed depth of $X=40$ cm.}
\label{fig3}
\end{figure}

In order to study the effect of the variation of the aspect ratio $B$, spacers were used corresponding to cavities of  $Y = \{2, 3, 4.5, 6.5, 10, 14, 20 \} $~cm and a fixed depth $X = 40$~cm for all cavities, that is, with values of the area $A = \{80, 120, 180, 260, 400, 560, 800\} $~cm$^2$ (one decade of variability). We choose $B>1$ in order to observe the consequences of the packing process in narrow cavities, i.e., the tendency for the trivial one-dimensional configuration. Figure~\ref{fig1} also belongs to this set of cavities corresponding to the tight packing of the rod in the widest cavity. Figure~\ref{fig3} shows some typical configurations obtained with the other cavities with $X=40$~cm.

The configurations which are shown in Fig.~\ref{fig3} present some similarity with those in Fig.~\ref{fig2}: for wider cavities there are loops that do not reach the inner boundaries of the cell, while for intermediary cavities the fraction of loops that touches the lateral walls increases. For this last type of cavity, as the aspect ratio grows, there is a progressive tendency of the rod to form loops restricted to the region close to the injection channel. The narrow patterns as shown in Fig.~\ref{fig3} were already observed for a more complicated system where the cavity is populated with a large number of pins \cite{gomes10}, but in the present case the number of loops is severely reduced to zero for $Y\approx 4 \zeta=2$ cm.
\newpage

\subsection{Cavities with $m$ circular discs}

\begin{figure}[!ht]
\centering
\includegraphics[height=0.30\linewidth]{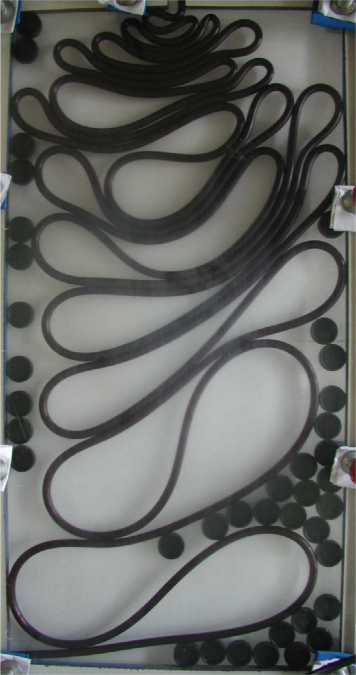}
\includegraphics[height=0.30\linewidth]{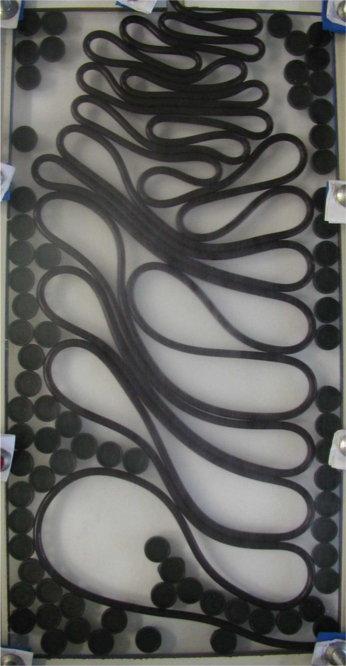}
\includegraphics[height=0.30\linewidth]{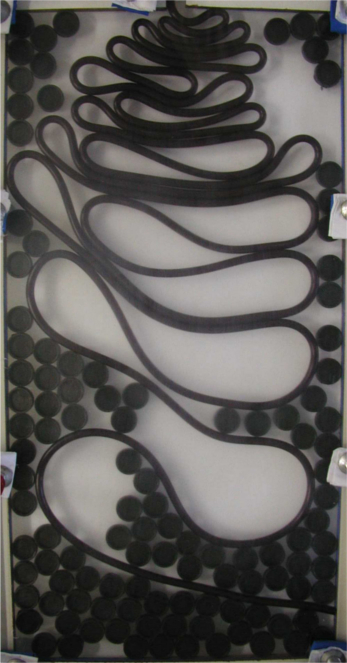}
\includegraphics[height=0.30\linewidth]{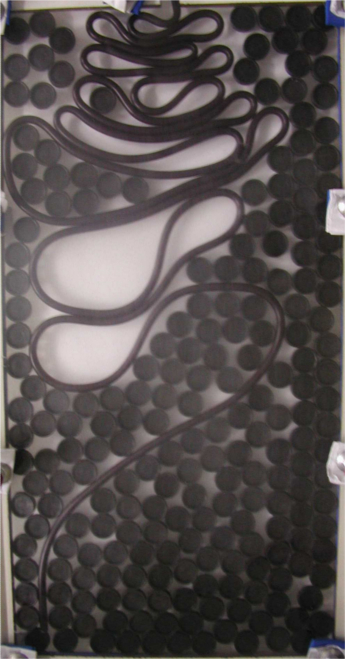}
\includegraphics[height=0.30\linewidth]{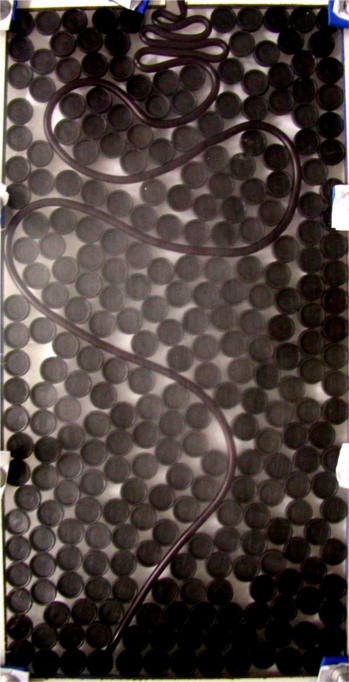}
\includegraphics[height=0.30\linewidth]{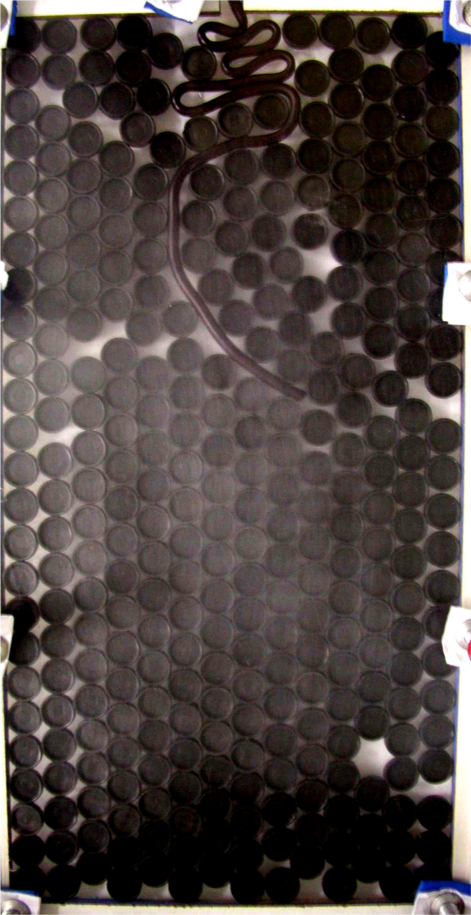}
\caption{Tight packing of a rod in cavities of $X = 40$ cm and $Y = 20$ cm in the presence of $m=\{38, 76, 114, 214, 283, 318\}$ (from left to right) circular discs.}
\label{fig4}
\end{figure}

Besides its intrinsic theoretical interest, the injection of a rod in cavities with discs could be of utility to studies of polymer packing in natural environments, because other structures compete for the occupation of the space. As an example, DNA molecules condense in chromosomes in the presence of proteins called histones. In our study the cavity of 20 cm $\times$ 40 cm was filled with $m$ circular discs of height $h=0.4$ cm and diameter $\phi = 1.6$ cm  (Fig.~\ref{fig4}).  Those rigid free discs are initially shaken in such a way that most discs do not reach the cavity boundaries, so that the rod interacts with the discs through the injection. 

Considering the random close packing fraction for circular discs as $p = 0.84$~\cite{weaire08}, it was estimated that an available area of $A=8$ cm$^2$ to the rod corresponds to about $m=331$ discs. It was taken several values $m =$  \{19, 38, 57, 76, 95, 114, 171, 214, 267, 283, 299, 318, 327, 331\} for the number of discs in the cavity in order to have two decades of area variability. Figure~\ref{fig4} shows some configurations for a rod in the rectangular cavity with $m$ circular rigid free discs. 

For small number of discs, the configurations are not different from the patterns found with cavities without discs (Fig.~\ref{fig2} and Fig.~\ref{fig3}): the discs are pushed to the boundaries of the cavity and the rod bends in the available area. For an intermediary number of discs, the higher concentration of discs influences the loops, in such a way that its formation is progressively inhibited and the remaining loops stay close to the injection channel. For a very high number of discs the initial configurations of discs (without the rod) present an almost positional and orientational ordered state, and the rod disposes itself in the spatial defects of the packing of discs. All pictures shown in Fig.~\ref{fig4} were chosen among a set of 10 for every configuration due to its quality and clarity. 

The configurations shown in Fig.~\ref{fig4} are very similar to those found  for crumpled wires in rectangular cavities in the presence of a large number $m$ of fixed pins (columnar phase~\cite{gomes10}). In particular, if the injection were performed through two opposite channels a columnar structure would be expected, with a set of few loops occupying the regions near the injection channels. We observed the formation of loops irrespective of the number of discs.


\section{Results and discussion}
\label{secRD}

With the increase of the injected length $l$ the number of loops $n$ increases in such a way that it is natural to expect a dependence $n=n(l)$ as presented in the literature \cite{donato02,donato03,stoop08,cunha09}. However, what is the element which limits the total length $L \equiv l_{max}$ of a rod that can be injected inside two-dimensional cavities? The intuitive answer for this question is the elasticity of the rod and its features of self-avoidance. These elements depend both on how the loops are located as well as on its size distribution in the cavity. With this argument, it can be expected that in the tight packing the total length of the rod is dependent on the total number of loops inserted, $N \equiv n_{max}$:
\begin{equation}
L-X=f(N,A).
\label{eq1}
\end{equation}
As an example, a rigid wire will not bend and its length will be equal to the depth $X$ of the cavity which corresponds to $f(0,A)=0$. For flexible rods there are more loops and more length is injected in the tight packing. Here, the total quantity of loops is seen as a dimensionless measure of the elasto-plastic features of the rod. 

The mass or length of wire that can be injected into a two-dimensional cavity depends also on the size of the cavity. We study the influence of the available area in rectangular cavities by three different ways: ({\it i}) by controlling the total area of the cavity keeping fixed the aspect ratio; ({\it ii}) by controlling the width of the cavity keeping fixed its depth; and ({\it iii}) by inserting a number of circular discs. The first and second settings can be seen as a study of the finite size effect for this system and can be used to establish a mass-size relationship \cite{maycon08}, while the latter corresponds to a model of an elastic one-dimensional object embedded in an environment partially occupied by a competing ``fluid''.  


\subsection{Cavities without discs}
\label{secCWO}

Our first proposal to the function $f(N,A)$ in Eq.~\ref{eq1} was within the format $f(N,A) \sim N^{\alpha} A^{\beta}$. Although the exponent $\alpha$ was found robust over all experiments, $\beta$ assumed two different values: $\beta_B = 0.51 \pm 0.01$ for the set of data of cavities with fixed aspect ratio $B$, and $\beta_X = 1.02 \pm 0.02$ for the set of cavities with fixed depth $X$. Our conclusion from these different values is that the length $L-X$ in the tight packing is sensitive only to the width of the cavity $Y$, so that $Y=\sqrt{A/2}$ for $B=2$, and $Y=A/40$ for $X=40$ cm. Therefore
\begin{equation}
L-X=kN^\alpha Y
\label{eq2}
\end{equation}
is a valid expression for all our experiments performed in rectangular cavities without discs. Please observe that Eq.~\ref{eq2} separates the contribution of the depth of the cavity $X$ and its width $Y$. The depth is the minimum length to be inserted into the cavity in tight packing, and the rod bends itself from one side to another inside the cavity, which gives a length proportional to the available width. Besides, we can see that the formation of loops is inhibited, $N = 0$, in both datasets (see Figs.~\ref{fig2} and~\ref{fig3}) when $Y = 2$ cm with no obvious association with the area of the cavity.

\begin{figure}[!ht]
\centering
\includegraphics[width=0.6\linewidth]{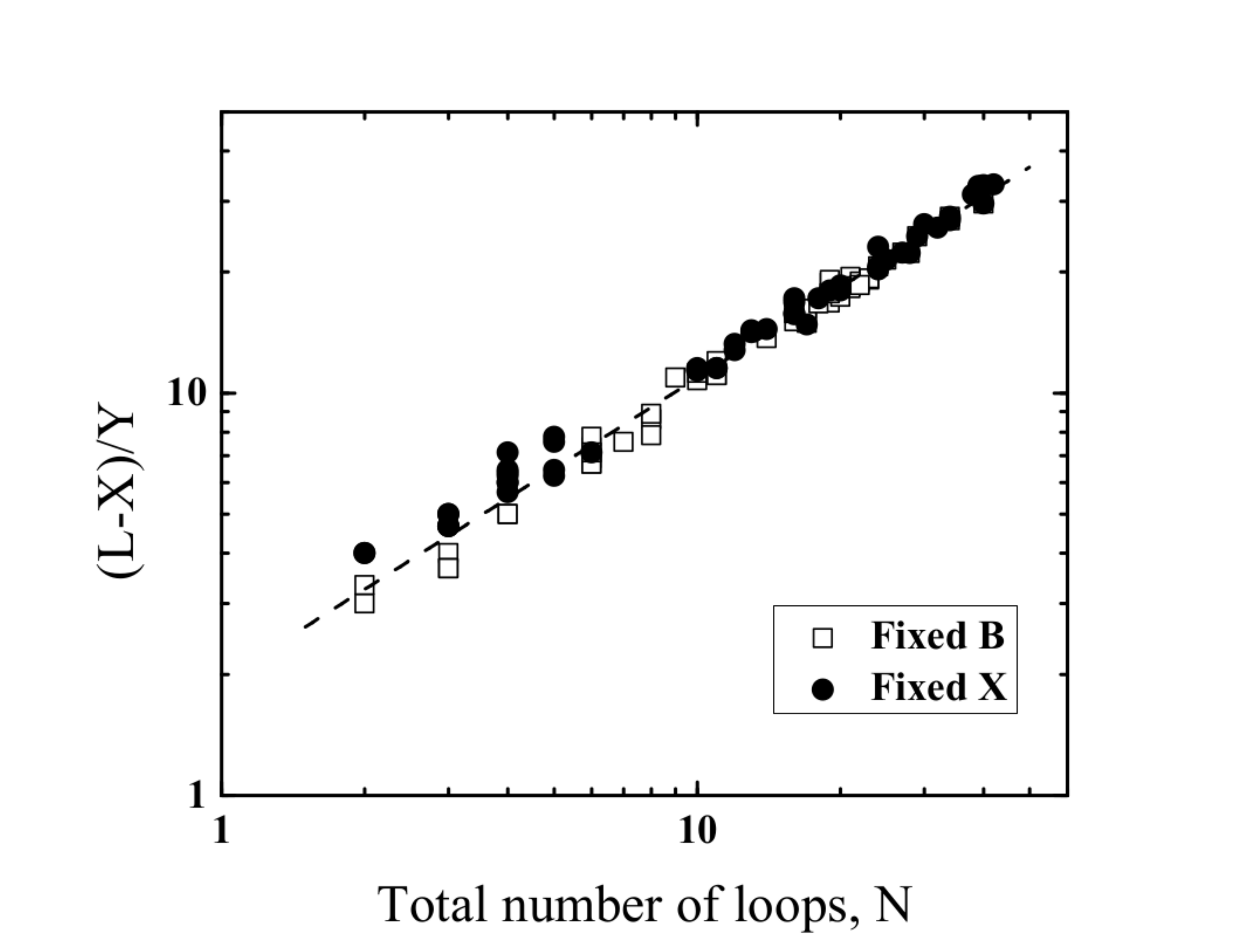}
\caption{The dependence of the $(L-X)/Y$ with the total number of loops $N$. The data for cavities with $B=2$ (open squares) and with $X=40$ cm (full circles) give a best power law fit with exponent $\alpha = 0.75$.}
\label{fig5}
\end{figure}

In order to obtain the exponent $\alpha$, the ratio $(L-X)/Y$ as a function of the total number of loops is shown in Fig.~\ref{fig5} for all data obtained from cavities without discs. Separately, the exponents found are $\alpha_B = (0.76 \pm 0.02)$ and  $\alpha_X = (0.73 \pm 0.02)$. The best fit of all data is shown in Fig.~\ref{fig5} by a dashed line whose corresponding exponent is $\alpha = (0.75 \pm 0.02)$. The value found for the proportionality constant in Eq.~\ref{eq2} is $k = (1.94 \pm 0.06)$. With these values in Eq.~\ref{eq2} all experimental data for the total length of a rod in tight packing can be obtained within an accuracy of about 5\% or less by counting the total number of loops.

The set $X$, $Y$, $L$ and $N$ are variables that, together, behave as stated in Eq.~\ref{eq2}. However, the Eq.~\ref{eq2} is not able to predict what will happen with $L$ when the parameters $X$ and $Y$ of the cavity change, because the total number of loops $N$ must change as well. From this argument, it is not correct to state that $L-X$ is independent of the depth $X$ because its dependence is implicit in the total number of loops $N$,  and this quantity in the present study is taken as an independent variable. From these considerations we treat Eq.~\ref{eq2} as a equation of state for the tight packing of $N$ loops in a rectangular cavity of depth $X$ and width $Y$.


\subsection{Cavities with free discs}
\label{secCO}

For a rectangular cavity of $Y=20$~cm width and $X=40$~cm depth occupied by $m$ free discs the situation is much more complex, because the cavity is not-simply-connected, the topology depends on the number of discs, and, in addition, there is the possibility of formation of clusters of discs tightly packed. Each free disc occupies an area of $\pi \phi^2/4 = 2.01$~cm$^2$, which corresponds to an available area for the rod of $A(m) = (800 - 2.01 m)$~cm$^2$. This indicates that the free area of the cavity is equal to the area of 398 discs. There is a lower quantity of discs that can be packed inside the cavity because in the close packing of discs there is a vacant space between the discs. We estimate  that a total quantity of $M = 361$~discs fits into the cavity through the assumption of a triangular lattice of discs (maximal packing fraction of $\pi/(2\sqrt{3}) = 0.907$).

\begin{figure}[!ht]
\centering
\includegraphics[width=0.6\linewidth]{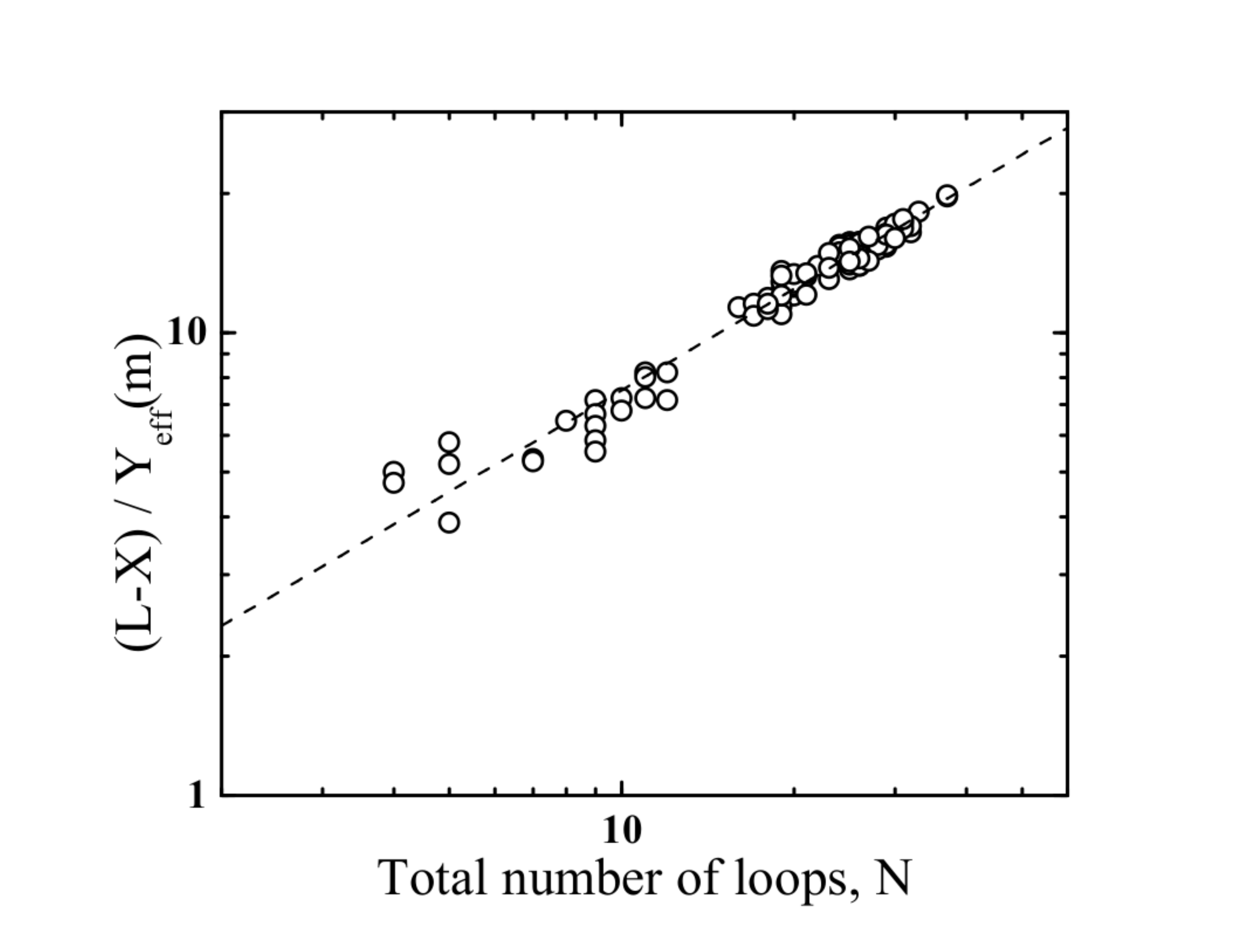}
\caption{The dependence of $(L-X) / Y_{eff}(m)$ with the total number of loops is a power law with exponent $\alpha_m = 0.73$, for $m$ in the interval [19, 283] (see subsection \ref{secCO} for details).}
\label{fig6}
\end{figure}

In spite of the complexities introduced in this type of packing, the experimental data is still described as outlined in the previous subsection: the length of the rod in the tight packing still obey the ansatz $f(N,A) \sim N^{\alpha} A^{\beta}$ with a constant exponent, for $m \leq 283$. The best fit for this range gives $\beta_m = 0.49 \pm 0.02$, which is the same exponent found for cavities with a given aspect ratio $B$. Here we should emphasize that the shape of the available area of the cavity is intricate in the present experiment. Defining an effective width $Y_{eff}(m) = \sqrt{A(m)}$, we plot in Fig.~\ref{fig6} the dependence of $(L-X)/Y_{eff}(m)$ on the total number of loops, for $m \leq 283$ discs.  The best power law fit is represented in Fig.~\ref{fig6} by a dashed line and gives the exponent $\alpha_m = 0.73 \pm 0.02$, which is equal to that found for cavities without discs.

In the previous subsection, the exponents in Eq.~\ref{eq2} were found constant among a set of several areas and shapes, although the cavities were always rectangular. In the present subsection the rod can spread throughout a more complicated space built by its interaction with the rigid discs and the exponents are found the same. Altogether the results suggest an independence on the shape of the rigid cavity, and stress the applicability of Eq.~\ref{eq2} in other studies of tight packing of rods. 

\begin{figure}[!ht]
\centering
\includegraphics[width=0.6\linewidth]{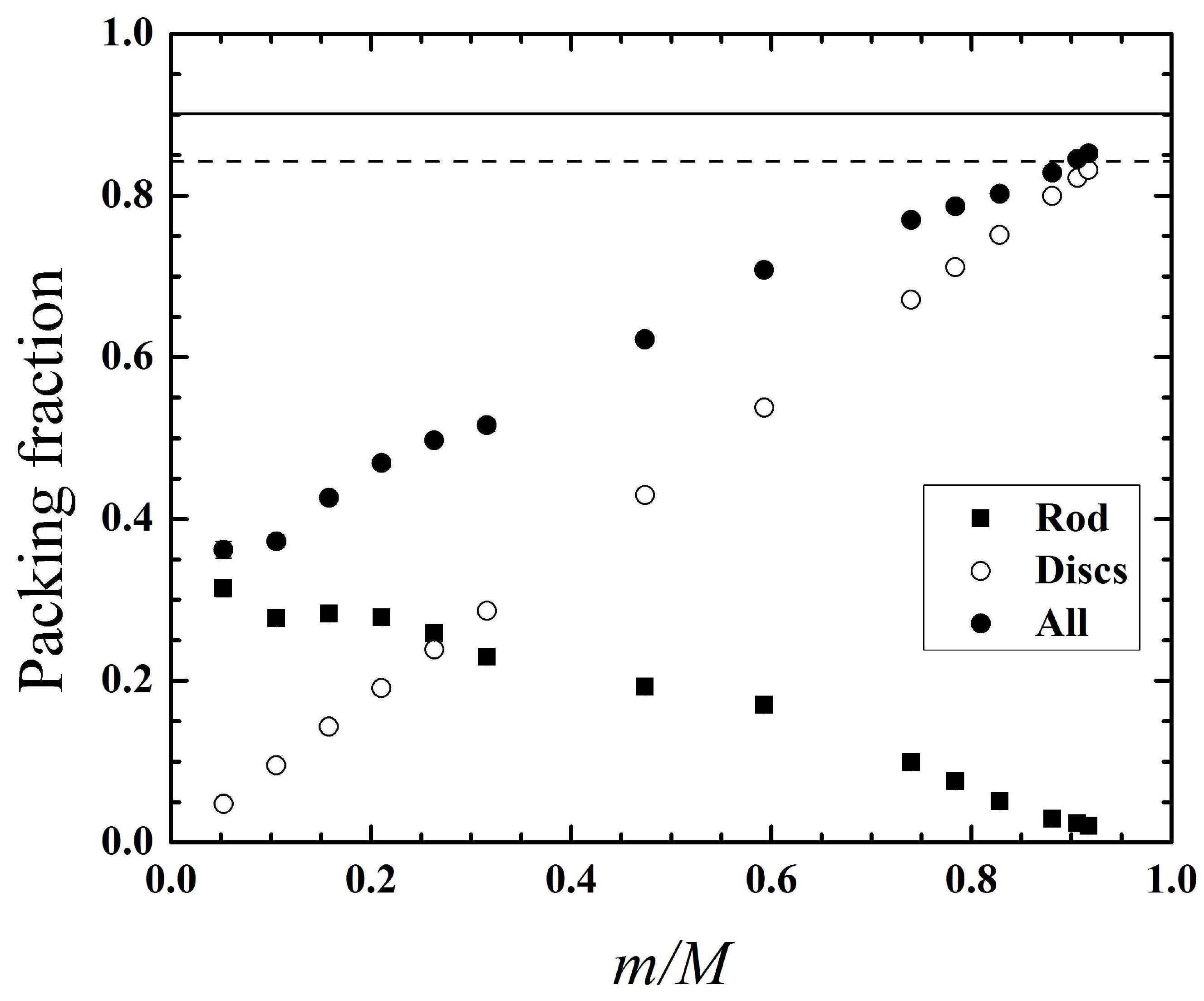}
\caption{The packing fraction of the rod (square), of the discs (open circles), and of rod + discs inside the cavity (close circles) as a function of the fraction of the number of discs that cover the cavity in a triangular lattice. The dashed line shows the random  close packing fraction for discs ($p=0.842$) and the continuous line shows the best ordered packing fraction of discs in two-dimensions ($p=0.907$).}
\label{fig7}
\end{figure}

In order to appreciate the connection of the problem studied here with the classical problem of two-dimensional packing of discs, we show in Fig.~\ref{fig7} a plot of the coverage fraction of the cavity for rod and discs separately, as well the total rod+discs coverage fraction, all them as a function of the normalized number of discs, $m/M$. As a guide to the eyes, we show in Fig.~\ref{fig7} the random close packing fraction of discs ($p=0.842$) and the maximal packing fraction of discs in two-dimensions ($p = 0.907$) obtained with a triangular array of discs in contact. The error bars are of the size of the symbols or smaller.

The packing fraction for the rod, $p_{rod} = L \zeta / A$, is a decreasing function of the number of discs.  However, the fluctuations shown in Fig.~\ref{fig7} are an indication of the complexity of the interaction rod-discs. The packing fraction for the discs, $p_{discs} = m (2.01/800)$, is shown in Fig.~\ref{fig7} for reference purposes. It can be visualized that for $m/M > 0.3$, that is, $m>107$ discs, the total area covered by the rod is smaller than the area covered by the discs. Interestingly, the value of the coverage fraction for rod+discs in higher values of $m/M$ exceeds the random close packing fraction for discs.


\subsection{Geometric models}
\label{secSM}

In an extremely compact two-dimensional packing of a rod, the whole space of the cavity is occupied, and the area covered by the rod is then $ \zeta L = XY$. In this case the rod bends itself over $\left( X / \zeta \right)$ parallel layers of pieces of rod with length equal to the width $Y$ of the cavity. The total number of loops is given by $N = (X / \zeta - 1)$ and the length in the tight packing is $L = (N+1)Y$. This simple model is able to give us insights about the proportionality between $L$ and $Y$ in Eq.~\ref{eq2}.

\begin{figure}[!th]
\centering
\subfigure[\label{fig8a}]{\includegraphics[width=0.2\linewidth]{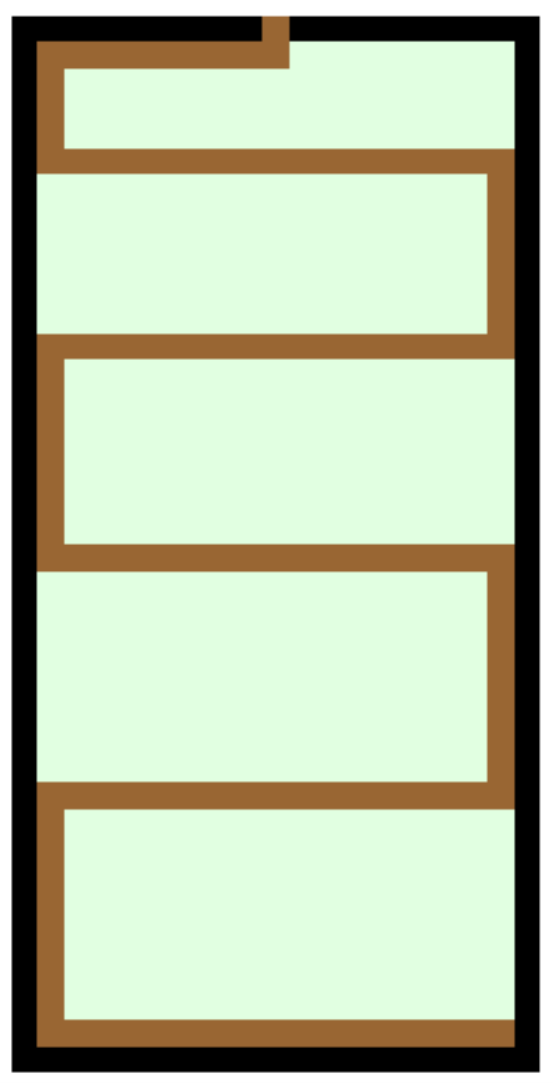}}
\hspace{30pt}
\subfigure[\label{fig8b}]{\includegraphics[width=0.2\linewidth]{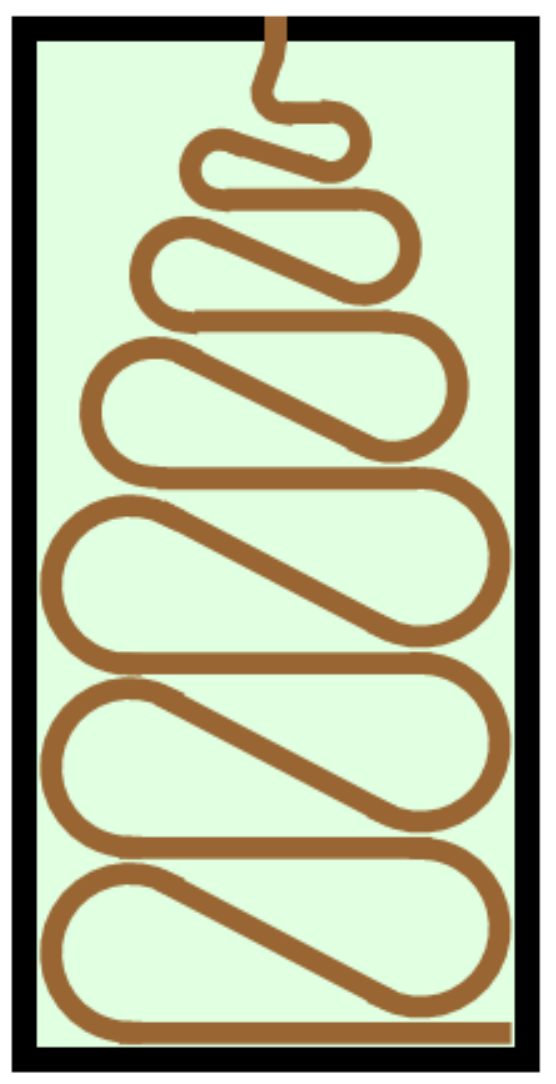}}
\caption{Geometric models for the packing of a rod in a rectangular cavity: \subref{fig8a} a  model based on straight lines and \subref{fig8b} a model based on circles and straight lines.  See subsection \ref{secSM} for details.}
\label{fig9}
\end{figure}

Figure \ref{fig8a} shows a slightly more realistic model which considers layers of rod spaced by a set of distances $\{b_i\}$, identified as the width of the loops. The loops are defined by the rod and the longitudinal line of symmetry of the cavity in such a way that the perimeter of each loop is then $p_i=\left(Y+b_i\right)$ and the total length inside the cavity is the sum $L=Y/2+\sum_{i=1}^{N} p_i = \left( N +1/2\right) Y + X$ for any distribution of values $\{b_i\}$ since $\sum_i b_i = X$. This model treats the number of loops $N$ as a packing independent parameter and shows the importance of subtracting the depth $X$ of the cavity.

Inside the cavity, the loops change their height $h_i$ along the $Y$-edge as well as their width  $b_i$ depending on its position. If we consider the loops as two-dimensional objects with a fixed shape, a relationship $h_i \sim b_i$ is expected (Fig.~\ref{fig8b}). Such gradient of heights must imply a conical structure as it is observed experimentally for the loops near the injection channel. That structure is supposed to have a fractal scaling \cite{donato02} that supports the expected asymptotic behaviour $(L-X)/Y \sim N^\alpha$.

The two-dimensional character of the tight packing studied here must present proportionality between the area covered by the rod, $\zeta L$, and the total area of the cavity, $XY$. By comparing this with Eq.~\ref{eq2} it is suggested that assymptotically $X \sim N^{3/4}$. Thus, if the cable is inserted from the middle of the larger side of the box, it would create fewer loops of larger sizes and consequently with a lower elastic energy. Furthermore, this relationship can be thought as a measure of the diffusion of the wire throughout the cavity. Interestingly, the number $3/4$ is found as the exponent which the end-to-end distance scales with the number of steps for the self-avoiding walk in two-dimensions~\cite{gennes79} which shares with the present system the features of self exclusion and one-dimensional topology.


\section{Conclusions}
\label{secCP}

In this paper a basic problem in two-dimensional physics is studied, namely the packing of a rod, a fibre, or a wire into a rectangular cavity. Although the problem studied here have an intrinsic interest, it could have some additional interest in a number of technological devices and natural processes aiming at accommodating, manipulating, and storing an object with one-dimensional topology in confined environments~\cite{purohit05,smith11,sakaue06,gennes79,Schmaljohann06,solovev12}. The optimal injection of a long rod of length $L$ into a finite two-dimensional cavity gives origin to a cascade of $N$ loops which are, in general, heterogeneously distributed along the cavity~\cite{donato02,stoop08,maycon08,donato03,donato07}. Here, the experimental study of this process is extensively investigated in 27 types of cavities of several sizes and topologies. In many of the packing processes examined in the present paper the cavity is partially occupied with mono-disperse discs in a geometry that is reminiscent of two important physical problems: the packing of a polymer-like structure in a confined geometry~\cite{sakaue06,gennes79} and the classical problem of packing of discs~\cite{weaire08}. In spite of the diversity of cavities used in the experiments, it is shown that a single expression or equation of state (Eq.~\ref{eq2}) is able to properly correlate the macroscopic variables of the rod, $L$ and $N$, with the dimensions and characteristic topologies of the cavities. We conjecture that this equation of state describes the packing of rods in the jamming limit for any two-dimensional cavity, irrespective of the shape of its contour, its topology, and also its intrinsic curvature, that is, including positive and negative curvatures as well.


\ack

We acknowledge the anonymous referees for the suggestions and critical reading of the text.  This work was supported by Brazilian Agencies CNPq (Conselho Nacional de Desenvolvimento Cient\'ifico e Tecnol\'ogico) and PRONEX (Programa de N\'ucleos de Excel\^encia).

\section*{References}

\end{document}